 \documentclass[final,3p,times]{elsarticle}
\usepackage[applemac]{inputenc}
\usepackage[english]{babel}
\usepackage{listings} 
\usepackage{fancyhdr}
\usepackage{graphicx}
\usepackage[caption=false]{subfig}
\usepackage[fleqn]{amsmath}
\usepackage{mathtools}
\usepackage{amsfonts}
\usepackage{acronym}
\usepackage{amssymb}
\usepackage{color}
\usepackage{float}
\usepackage{subfloat}
\usepackage{hyperref}
\usepackage{enumerate}
\usepackage{pgfkeys}
\usepackage{rotating}
\usepackage{rotfloat}
\usepackage{tabulary}
\usepackage{natbib}
\usepackage{epstopdf}
\usepackage{stmaryrd}
\usepackage{setspace}
\usepackage{gensymb}
\usepackage{verbatim}
\usepackage{tikz}
\usetikzlibrary{arrows,shapes,calc}
\usetikzlibrary{decorations.pathreplacing}
\usetikzlibrary{intersections}
\usepackage{nomencl}

\makenomenclature
\newcommand\Rey{\mbox{\textit{Re}}}  
\newcommand{\diff}{\mathrm{d}}
 
\journal{Computers \& Fluids}

\begin{document}
\begin{frontmatter}

 \title{A high-fidelity solver for turbulent compressible flows on unstructured meshes}
\author{Davide Modesti~$^\ast$ and Sergio Pirozzoli}
\ead{davide.modesti@uniroma1.it}
\cortext[cor1]{Corresponding author. Tel.  +39-06-44585202, Fax +39-06-44585250}
\address{Sapienza Universit\`a di Roma, Dipartimento di Ingegneria Meccanica e Aerospaziale, via Eudossiana 18, 00184 Roma, Italy}

\begin{abstract}
We develop a high-fidelity numerical solver for the compressible Navier-Stokes equations,
with the main aim of highlighting the predictive capabilities of low-diffusive numerics for
flows in complex geometries. The space discretization of the convective terms in the
Navier-Stokes equations relies on a robust energy-preserving numerical flux, 
and numerical diffusion inherited from the AUSM scheme is added limited to the vicinity of shock waves, 
or wherever spurious numerical oscillations are sensed.
The solver is capable of conserving the total kinetic energy in the inviscid limit, and it bears sensibly less
numerical diffusion than typical industrial solvers, with incurred greater predictive power,
as demonstrated through a series of test cases including DNS, LES and URANS of turbulent flows.
Simplicity of implementation in existing popular solvers such as OpenFOAM$^{\textregistered}$ is also highlighted.
\end{abstract}
\end{frontmatter}

\section{Introduction}

Computational fluid dynamics (CFD) has become a common tool
for the prediction of flows of engineering interest. 
Since the pioneering works of~\citet{orszag_72,kim_87}, which first showed
the potential of computers for high-fidelity prediction of turbulent flows,
many studies have appeared in which CFD has been used to tackle fundamental topics in 
turbulence research~\citep{schlatter_10,sillero_13,bernardini_14},
and to solve flows of industrial interest~\citep{kim_99,iaccarino_01,mahesh_06,bernardini_16}.
Although CFD is currently used 
with good degree of success in the routine industrial design process, 
a large disparity between the accuracy of algorithms used in commercial
flow solvers and in academia is still evident.
Spectral methods~\citep{hussaini_87}, high-order finite difference (FD) 
methods~\citep{lele_92}, discretely energy-preserving schemes~\citep{harlow_65,orlandi_12}, 
and accurate explicit time integration~\citep{jameson_81,suresh_97}
are common features of many academic flow solvers.
Accurate techniques are also available to capture shock waves in compressible
flow, which include the essentially-non-oscillatory schemes and their weighted
counterpart, or hybrid schemes \citep{pirozzoli_02,pirozzoli_11,hickel_14}.
On the other hand, most commercial flow solvers rely 
on first/second order unstructured finite volume (FV) discretizations,
in which the nonlinear terms are typically stabilized through upwinding,
and time is advanced through implicit 
segregated algorithms \citep{patankar_72,ferziger_12}.
In the case of compressible flows,
shock-capturing capability is frequently achieved 
through sturdy but outdated total-variation-diminishing (TVD) schemes, or rougher.
A common feature of most commercial flow solvers is 
the use of severely diffusive numerical algorithms,
which may negatively impact the prediction of unsteady turbulent flows,
especially in large-eddy simulation (LES)~\citep{mittal_97}.
Although high-accuracy, energy-consistent discretizations can also be applied
to unstructured meshes of industrial relevance~\citep{nicolaides_97,ducros_00,perot_00}, 
it appears that the approach has not been incorporated in solvers of common use.
The main aim of this work is trying to bridge this gap, by introducing high-fidelity low-diffusive 
numerical schemes of academic use 
into existing unstructured flow solvers, with the eventual intent of
achieving more accurate prediction of turbulent flows of industrial interest, 
possibly with little computational overhead.
For illustrative purposes, we consider as baseline solver
the open-source library OpenFOAM$^{\textregistered}$~\citep{weller_98}, 
which is released
under the General Public Licence (GPL), and which has experienced large diffusion
in the recent years.
The baseline distribution of OpenFOAM$^{\textregistered}$ comes with
several compressible flow solvers, of which the most widely used is 
\textit{rhoCentralFoam}, relying on full discretization of the convective fluxes
through the central TVD scheme of \citet{kurganov_00}.

Some attention has been recently devoted to
modification of the standard OpenFOAM$^{\textregistered}$ algorithms with the goal of reducing
their numerical diffusion~\citep{vuorinen_12,vuorinen_14}.
For instance, \citet{vuorinen_12,vuorinen_14} have introduced a scale-selective 
mixed central/upwind discretization 
which is particularly beneficial for LES, especially when coupled
with low-diffusion Runge-Kutta time integration.
Although the approach limits the amount of numerical diffusion,
discrete conservation of total kinetic energy in the inviscid limit is not guaranteed.
\citet{shen_14,shen_16} developed an implicit compressible solver for OpenFOAM$^{\textregistered}$
relying on the AUSM scheme~\citep{liou_93} and found similar performances as \textit{rhoCentralFoam}. 
\citet{cerminara_16} developed a compressible multi-phase solver for OpenFOAM$^{\textregistered}$ based 
on the PIMPLE algorithm~\citep{ferziger_12} for the simulation of volcanic ash plumes, which 
is considerably less diffusive than \textit{rhoCentralFoam}.
Hence, it appears that the OpenFOAM$^{\textregistered}$ community is concerned about numerical diffusion, and 
some effort is being devoted to trying to minimize it, both for incompressible and compressible flows. 
Herein we describe an algorithm for the numerical solution of the compressible 
Navier-Stokes equations which allows to discretely preserve the total flow kinetic
energy from convection in the inviscid limit on Cartesian meshes~\citep{pirozzoli_10},
and to maintain good conservation properties also on unstructured triangular meshes 
through localized augmentation of the numerical flux with the AUSM pressure diffusive flux.
Shock-capturing capability is further obtained through localized use of
the full AUSM diffusive flux, wherever shocks are sensed.  
The full algorithm is illustrated in detail in Section~\ref{sec:numerics}, 
and the results of several numerical tests reported in Section~\ref{sec:results}.
Concluding remarks are given in Section~\ref{sec:conclusions}

\section{Numerical algorithm} \label{sec:numerics}

 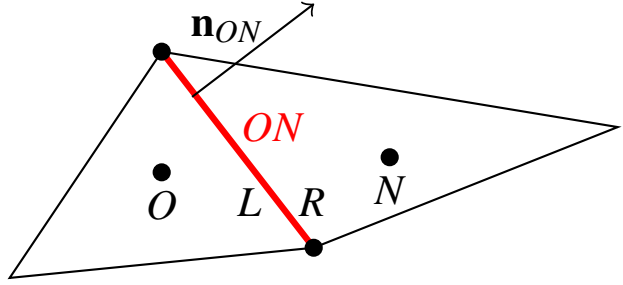
\begin{figure}
 \centering
\scalebox{2}{
 \begin{tikzpicture}
 \coordinate (A) at (7,1);
 \coordinate (E) at (7.78,0.0);
 \coordinate (B) at (6,-0.5);
 \coordinate (C) at (8,-0.3);
 \coordinate (D) at (10,0.5);
 \coordinate (C1) at (7,0.2);
 \coordinate (C2) at (8.5,0.3);
 \coordinate (C3) at (7.2,0.7);
 \coordinate (n) at (7.8,0.9);
 \coordinate (AB) at (7.4,0.5);
 \draw (A)--(B)--(C)--cycle;
 \draw (A)--(C)--(D)--cycle;
 \draw[line width=1.2pt,color=red] (A)--(C);
 \draw[->] (C3) -- ($(A)!1.05cm!70:(C)$) ;
 \draw[fill=black] (n) node [above=0.7em,left]{\footnotesize$\mathbf{n}_{ON}$};
 \draw[fill=black] (C1) circle(0.15em) node [below]{\footnotesize$O$};
 \draw[fill=black] (C2) circle(0.15em) node [below]{\footnotesize$N$};
 \draw[fill=black] (A) circle(0.15em) node [above]{\footnotesize$$};
 \draw[fill=black] (C) circle(0.15em) node [below]{\footnotesize$$};
 \draw[fill=blue] (E)  node [ left]{\footnotesize$L$};
 \draw[fill=blue] (E)  node [ right]{\footnotesize$R$};
 \draw[color=red] (AB) node [right]{\footnotesize$ON$};
 \end{tikzpicture}}
\caption{Computational cell for evaluation of the numerical flux.
        $ON$ denotes the interface between the owner cell $O$, and the 
        neighbouring cell $N$, and $\mathbf{n}_{ON}$ is the
        outer normal for $O$.
        $L$ and $R$ denote limit states at the two sides of the interface
        ($L \equiv O$, $R \equiv N$ in the case of first-order reconstruction).
        }
\label{fig:cell}%
\end{figure}

We consider the Navier-Stokes equations for a compressible ideal gas, integrated over
an arbitrary control volume $V$
\begin{equation}
\frac {\diff}{\diff t} \int_V {\bf u} \, {\diff V} + \sum_{i=1}^3 \int_{\partial V} \left( {\bf f}_i - {\bf f}^v_i \right) \, n_i \, {\diff S}= 0, \label{eq:NS}
\end{equation}
where ${\bf n}$ is the outward normal, and
\begin{equation}
\mathbf{u} = 
\begin{Bmatrix}
\rho \\
\rho u_i\\
\rho E\\
\end{Bmatrix},\quad
\mathbf{f}_i = 
\begin{Bmatrix}
\rho u_i\\
\rho u_iu_j+p\delta_{ij}\\
\rho u_iH\\
\end{Bmatrix},\quad
\mathbf{f}_i^v =
\begin{Bmatrix}
 0\\
\sigma_{ij}\\
\sigma_{ik}u_k-q_i\\
\end{Bmatrix},
\label{eq:fluxes}
\end{equation}
are the vector of conservative variables, and the associated Eulerian and viscous fluxes, respectively.
Here $\rho$ is the density, $u_i$ is the velocity component in the $i$-th coordinate direction,
$p$ is the thermodynamic pressure, $E=e + u^2/2$ is the total energy per unit mass,
$e = R T / (\gamma - 1)$ is the internal energy per unit mass, $H=E+p/\rho$ is the total enthalpy,
$R$ is the gas constant, $\gamma =c_p/c_v$ is the specific heat ratio,
$\sigma_{ij}$ is the viscous stress tensor, and $q_i$ is the heat flux vector.
 
The boundary Eulerian flux in Eqn.~\eqref{eq:NS} is approximated on a polyhedral cell $O$ 
(see Fig.~\ref{fig:cell} for illustration) as follows 
\begin{equation}
\frac{1}{V}\int_{\partial V}\mathbf{f}_in_i\mathrm{d}S\approx
\sum_{N} (\mathbf{f}_in_i)_{ON}\mathrm{\Delta}S_{ON}= 
\sum_{N} \hat{\mathbf{f}}_{ON}\mathrm{\Delta}S_{ON},
\end{equation}
where $\hat{\mathbf{f}}_{ON}$ is the numerical flux at the interface between
the cell and its neighbour $N$, 
$\mathrm{\Delta}S_{ON}$ is the interface area, and $\sum_{N}$ denotes summation on all cell faces.

As customary in the AUSM approach~\citep{liou_93}, we proceed by splitting the Eulerian flux in Eqn.~\eqref{eq:fluxes} into
a convective and a pressure contribution, namely
\begin{equation}
\mathbf{f}_i = \mathbf{f}_i+\mathbf{p}_i= 
\begin{Bmatrix}
\rho u_i\\
\rho u_iu_j\\
\rho u_iH\\
\end{Bmatrix}
+
\begin{Bmatrix}
0\\
p\delta_{ij}\\
0
\end{Bmatrix},
\label{eq:flux}
\end{equation}
whose associated numerical fluxes are cast as the sum of a central and a diffusive part,
\begin{equation}
\hat{\mathbf{f}}_{ON} = \hat{\mathbf{f}}_{ON}^{C}
+ \hat{\mathbf{f}}_{ON}^{D}, \quad 
\hat{\mathbf{p}}_{ON} = \hat{\mathbf{p}}_{ON}^{C}
   + \hat{\mathbf{p}}_{ON}^{D}.
\label{eq:num_flux}
\end{equation}
The central part of the convective flux is here evaluated as follows~\citep{pirozzoli_10}
\begin{equation}
\hat{\mathbf{f}}_{ON}^{C} = 1/8 \left(\rho_O+\rho_N\right)\left({u_n}_O+{u_n}_N\right)\left(\boldsymbol\varphi_O+\boldsymbol\varphi_N\right), 
\label{eq:ec_flux}
\end{equation}
where $\boldsymbol{\varphi}=\left( \rho, \rho u_i, \rho H \right)^T$,
and the pressure flux is evaluated through standard central interpolation,
\begin{equation}
\hat{\mathbf{p}}_{ON}^{C}=1/2 \left( \mathbf{p}_O+\mathbf{p}_N \right).
\end{equation}
Unlike straightforward central differencing, the numerical flux \eqref{eq:ec_flux} 
allows to discretely preserve the total kinetic energy of the flow from convection, 
with incurred strong nonlinear stability properties.
The above central numerical flux is in fact found to be stable in fully resolved simulations 
(DNS) on Cartesian or weakly distorted meshes~\citep{pirozzoli_10,pirozzoli_11b}.
However, in the case of practical engineering computations on unstructured meshes,
and certainly if shock waves are present,
some (possibly small) amount of numerical diffusion is necessary.
Hence, the diffusive fluxes in Eqn.~\eqref{eq:num_flux} should be locally activated
wherever resolution is lost.
To judge on the local smoothness of the numerical solution we rely on a classical 
shock sensor~\citep{ducros_99}
\begin{equation}
 \theta = \max{\left(\frac{-\nabla\cdot{u}}{\sqrt{{\nabla\cdot u}^2+{\nabla\times u}^2+u_{0}^2/L_0}},0\right)} \in[0,1], \quad \theta_{ON} = 1/2 \left( \theta_O+\theta_N \right),
\label{eq:sensor}
\end{equation}
where $u_0$ and $L_0$ are suitable velocity and length scales~\citep{pirozzoli_11}, defined 
such that $\theta \approx 0$ in smooth zones, and $\theta \approx 1$ in the presence of shocks.

\begin{table}
 \centering
 \begin{tabular}{cccc}
 \hline
 Mode & Intent & IC & IP \\
 \hline 
 A & Fully resolved smooth flows & 0 & 0 \\ 
 B & Unresolved smooth flows     & 0 & 1 \\ 
 C & Shocked flows               & 1 & 1 \\ 
 \hline
 \end{tabular}
 \caption{Modes of operation of the flow solver, with corresponding suggested values for the flags in Eqn.~\eqref{eq:difflux}.}
\label{tab:modes}
\end{table}

In the case of smooth flows (no shocks) we have found that additional numerical stability
with minimal accuracy penalty can be achieved by applying the artificial diffusion term 
to the pressure flux only, in amount proportional to $\theta_{ON}$.
Capturing shock waves further requires concurrent activation of the convective diffusive flux,
wherever $\theta_{ON}$ exceeds a suitable threshold (say $\theta^*$, here set to $0.05$, unless explicitly stated otherwise).
Hence, the diffusive numerical fluxes to be used in Eqn.~\eqref{eq:num_flux} may be synthetically expressed as follows
\begin{equation}
\hat{\mathbf{f}}_{ON}^{D} = 
\mathrm{IC} \, H (\theta_{ON} - \theta^*) \, \hat{\mathbf{f}}_{ON}^{AUSM}, \quad
\hat{\mathbf{p}}_{ON}^{D} = 
\mathrm{IP} \, \theta_{ON} \, \hat{\mathbf{p}}_{ON}^{AUSM},
\label{eq:difflux}
\end{equation}
where IC and IP are flags controlling the activation of the convective and pressure 
diffusive fluxes, $H$ indicates the Heaviside step function, and 
the artificial diffusion fluxes are borrowed from the AUSM scheme, 
as reported for convenience in Appendix~\ref{sec:appendix}. 
Suggested values for IC and IP are given in Tab.~\ref{tab:modes}, according to the type 
of numerical simulation to be carried out.

Discretization of the viscous fluxes relies on standard second-order approximations
for unstructured meshes~\citep{hirsch_07}, which is implemented through the
$\mathrm{fvc::laplacian()}$ primitive of OpenFOAM$^{\textregistered}$.
The resulting semi-discretized system of ordinary differential equations, say
$\mathrm{d}\mathbf{u}/\mathrm{d}t = \mathbf{R}(\mathbf{u})$, is 
advancement in time using a low-storage third-order, four-stage Runge-Kutta algorithm,
\begin{equation}
\mathbf{u}^{(\ell)} = \mathbf{u}^{(0)} + \alpha_{\ell} \Delta t \mathbf{R} (\mathbf{u}^{(\ell-1)}), \quad \ell=1,\ldots,4,
\label{eq:RK}
\end{equation}
where $\mathbf{u}^{(0)}=\mathbf{u}^{n}$,
$\mathbf{u}^{n+1}=\mathbf{u}^{(4)}$,
with $\alpha_1=1/4$, $\alpha_2=1/3$, $\alpha_3=1/2$, $\alpha_4=1$.

\section{Results} \label{sec:results}

We hereafter present a series of test cases representative of the three modes of operation listed in Tab.~\ref{tab:modes},
with the goal of testing the energy-preserving capabilities of the present solver, here referred to as \textit{rhoEnergyFoam},
and compare its performance with standard OpenFOAM$^{\textregistered}$ solvers.
Inviscid homogeneous isotropic turbulence and Taylor-Green flow
are used to quantify numerical diffusion.
DNS of supersonic channel flow is used to compared with data from an academic finite-difference solver.
RANS and DES of subsonic turbulent flow past a circular cylinder 
are performed to test the effectiveness of background numerical diffusion for smooth flows. 
The shock-capturing capabilities are further tested
using three classical flow cases, namely the inviscid supersonic flow past
a forward-facing step, the transonic flow past a RAE airfoil, and the transonic flow past the ONERA M6 wing.

\subsection{Decaying homogeneous isotropic turbulence}

\begin{figure}
 \begin{center}
  \includegraphics[width=6.0cm]{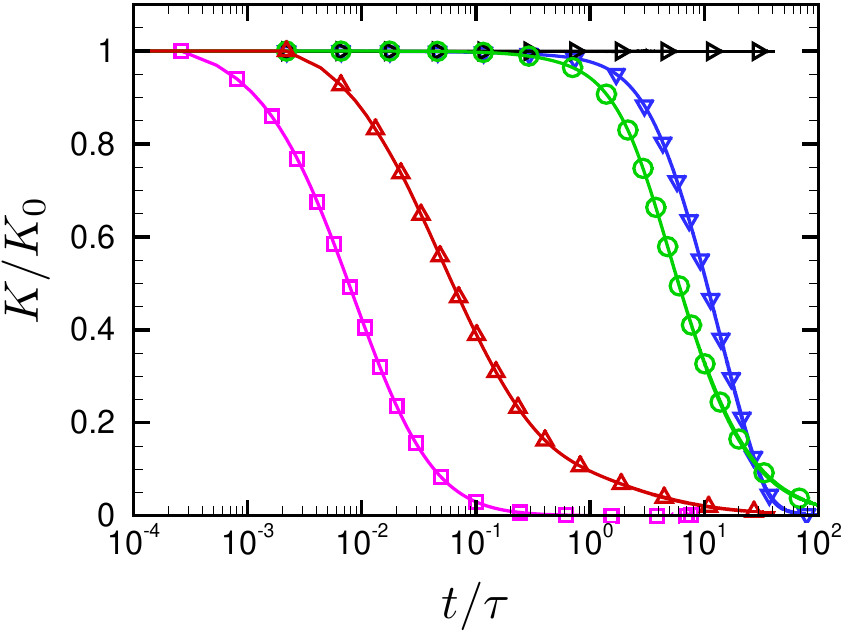}
  \caption{Decaying isotropic turbulence: time evolution of turbulence kinetic energy for \textit{rhoEnergyFoam} in Mode
           A (right triangles), Mode B (gradient symbols), Mode C with $\theta^*=0$ (deltas),
           and for \textit{dnsFOAM} (circles), \textit{rhoCentralFoam} (squares).
           $\tau$ is the eddy turnover time.}
  \label{fig:tke_hit}
 \end{center}
\end{figure}

In order to quantify the energy preservation properties of the present solver, numerical 
simulations of decaying homogeneous isotropic turbulence are carried out at zero physical viscosity. 
Random initial conditions are used with prescribed energy spectrum~\citep{blaisdell_91}, 
\begin{equation}
E(k) = 16 \sqrt{\frac 2{\pi}} \frac {u_0^2}{k_0} \left( \frac {k}{k_0} \right)^4 e^{-2 (k/k_0)^2},
\end{equation}
where $k_0 = 4$ is the most energetic mode, and $u_0$ is the initial r.m.s. velocity. 
The initial turbulent Mach number
is $M_{t0} = \sqrt{3} u_0/c_0 = 0.01$ ($c_0$ is the initial mean sound speed), and time is made nondimensional
with respect to the eddy turnover time $\tau = 2 \sqrt{3} / (k_0 M_{t0} c_0)$.
Numerical simulations are carried out on a $32^3$ Cartesian mesh with spacing $\Delta x$, and the time step 
$\Delta t$ is kept constant, 
corresponding to an initial Courant number $\mathrm{CFL}=\max{(u_0+c_0)} {\Delta t}/{\Delta x}=1$.
Figure~\ref{fig:tke_hit} shows the turbulence kinetic energy $K=1/2 \sum_i \overline{u_i u_i} V_i$,
as a function of time for \textit{rhoEnergyFoam} in the three modes of operation previously described.
Note that in the numerical experiments the threshold for activation of the convective diffusive fluxes is here momentarily set to zero,
to give a perception for the maximum possible amount of numerical diffusion in shock-capturing simulations.
For comparison purposes, results obtained with \textit{rhoCentralFoam} and with the 
OpenFOAM$^{\textregistered}$ incompressible DNS solver (\textit{dnsFoam}) are also shown.
It is clear that both baseline OpenFOAM$^{\textregistered}$ solvers are not capable of preserving the total kinetic energy,
because of the presence of numerical diffusion, which is higher in \textit{rhoCentralFoam}.
As expected, total kinetic energy is exactly preserved from \textit{rhoEnergyFoam} when operated in Mode A.
The addition of numerical diffusion to the pressure term (Mode B) causes some numerical 
diffusion, although still smaller than \textit{dnsFoam}, 
and most kinetic energy is in fact retained for one eddy turn-over time.
Operation in Mode C (with $\theta^*=0$) further increases numerical diffusion,
although the behavior is still sensibly better than \textit{rhoCentralFoam}. 

\subsection{Taylor-Green flow}

\begin{figure}
 \begin{center}
  \includegraphics[width=6.0cm]{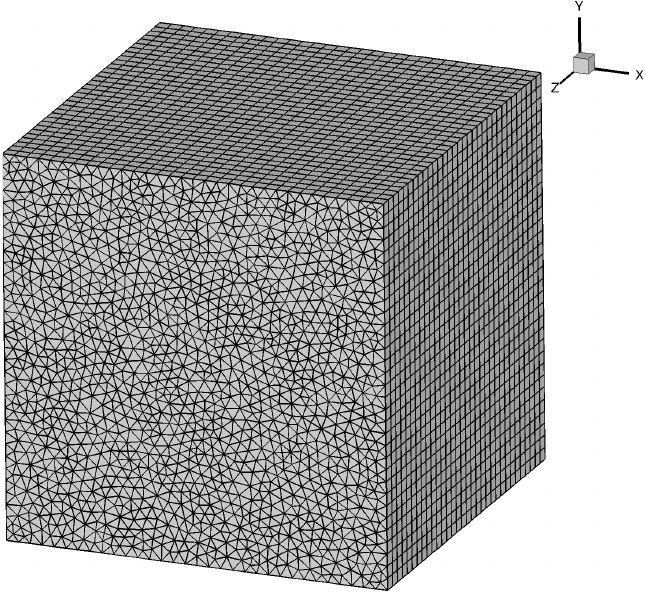}
  \caption{Unstructured mesh for numerical simulation of Taylor-Green flow.}
  \label{fig:mesh_tgv}
 \end{center}
\end{figure}
 
\begin{figure}
 \begin{center}
  (a)
  \includegraphics[width=5.0cm]{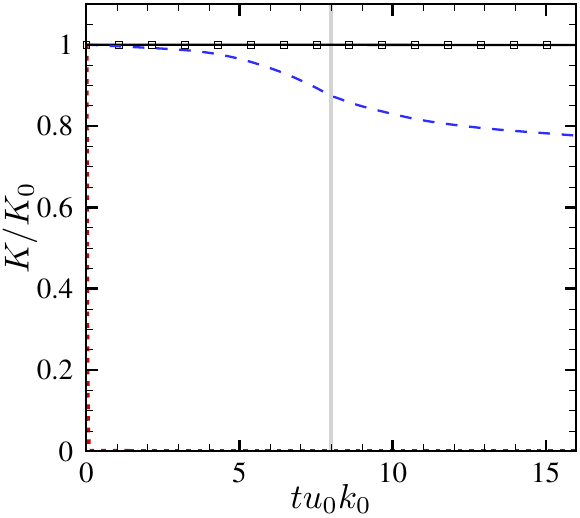} \hskip1.em
  (b)
  \includegraphics[width=5.0cm]{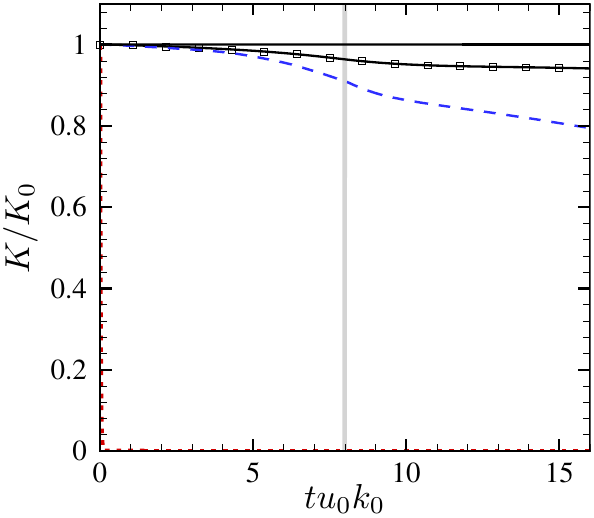}\\
  (c)
  \includegraphics[width=5.0cm]{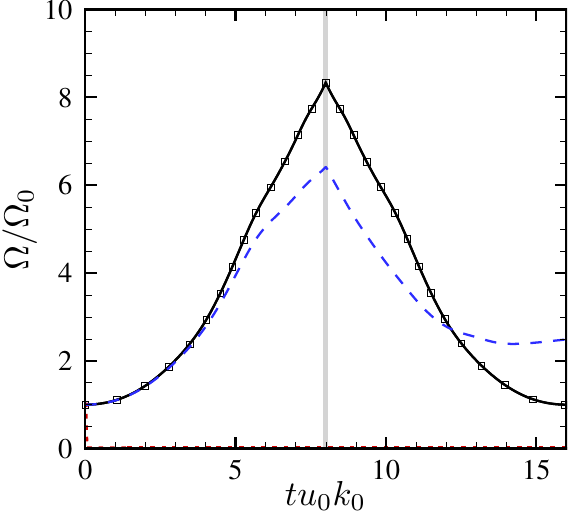} \hskip1.em
  (d)
  \includegraphics[width=5.0cm]{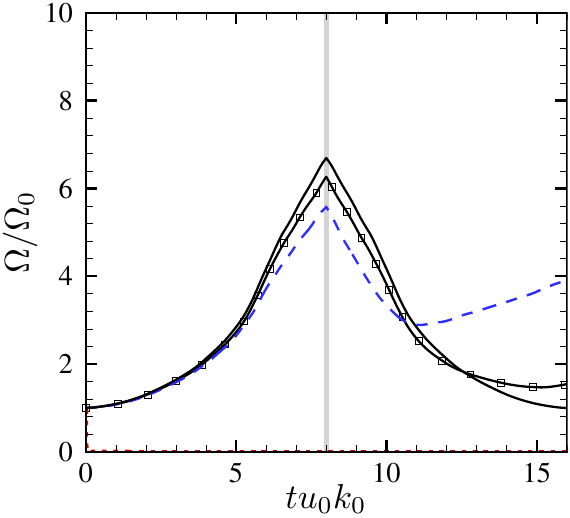} \\ \vskip1.em
  \caption{Time evolution of total kinetic energy (a-b), and enstrophy (c-d) 
    for Taylor-Green flow at $M_0=0.01$ 
    on Cartesian mesh (a-c) and on unstructured mesh (b-d).
    Data are shown for \textit{rhoEnergyFoam} in Mode A (solid lines), \textit{rhoEnergyFoam} in Mode B (solid lines with square symbols), \textit{dnsFoam} (dashed lines), and \textit{rhoCentralFoam} (dotted lines). 
    The vertical line indicates the time $t u_0 k_0=8$, at which velocity vectors are reversed.}
  \label{fig:tgv}
 \end{center}
\end{figure}

The energy-preserving properties of the solver are further tested for the case proposed by~\citet{duponcheel_08},
namely the time reversibility of the inviscid Taylor-Green flow.
The solution is computed in a $(2\pi)^3$ triply-periodic box, and initialized as follows
\begin{subequations}
\begin{align}
\rho &=\rho_0,\\
u&=u_0\sin{(k_0\,x)}\cos{(k_0\,y)}\cos{(k_0\,z)},\\
v&=u_0\cos{(k_0\,x)}\sin{(k_0\,y)}\cos{(k_0\,z)},\\
w&=0,\\
p& = p_0+u_0^2/16 [\cos{(2\,k_0\,z)}+2\,(\cos{(2\,k_0\,x)}+\cos{(2\,k_0\,y)})-2], 
\end{align}
\end{subequations}
where $k_0=1$ is the initial wavenumber, $u_0 = M_0 c_0$ is the reference velocity (here $M_0=0.01$),
and $c_0, p_0$, $T_0$, and $\rho_0$ are the reference
speed of sound, pressure, temperature and density.  
The Taylor-Green flow is widely studied as a model for turbulence formation from ordered
initial conditions, exhibiting rapid formation of small-scale structures with
incurred growth of vorticity.
This flow case is computed both on a Cartesian and an unstructured mesh. 
The Cartesian mesh has $32^3$ cells, whereas 
the unstructured mesh is obtained by extruding a two-dimensional mesh 
with triangular cells (see Fig.~\ref{fig:mesh_tgv}), hence including 85056 triangular prisms.
This setting guarantees exact geometrical
correspondence of the elements on opposite faces of the computational box, hence
periodicity can be exploited in all space directions.  
The solution is advanced in time up to time $t u_0 k_0 = 8$, at which
all velocity vectors are reversed, and then further
advanced in time up to $t u_0 k_0 = 16$. Based on the mathematical properties of the
Euler equations, the initial conditions should be exactly recovered~\citep{duponcheel_08}.
 
Numerical diffusion generally spoils time reversibility, as shown in Fig.~\ref{fig:tgv}
where we report the time evolution of turbulence kinetic energy and of 
the total enstrophy, defined as $\Omega = 1/2 \sum_i \overline{\omega_i \omega_i} V_i$.
The total kinetic energy (panels a, b) in fact shows monotonic decrease for 
\textit{dnsFoam} both on structured and unstructured meshes, 
and \textit{rhoCentralFoam} exhibits sudden dissipation of all kinetic energy, on
a time scale which is much less than unity (the lines are barely visible in the chosen representation).
On the other hand, kinetic energy
is almost perfectly retained by \textit{rhoEnergyFoam} when operated in Mode A, whereas
some effect of numerical diffusion is found in Mode B.
The total enstrophy computed on a Cartesian mesh (panel c) shows substantial growth 
up to time reversal, followed by corresponding decrease.
However, recovery of the initial condition is imperfect for \textit{dnsFoam},
and the maximum vorticity at the end of the simulation is higher than expected.
This odd behavior is associated with the flow randomization at the end of the 
forward run, which is not fully recovered in simulations contaminated by
numerical diffusion. On unstructured mesh (panel d) the behavior is similar, although 
the peak enstrophy is lower because of errors associated with mesh distortion.
Overall, this test shows that \textit{rhoEnergyFoam} retains good low-diffusive characteristics
also on unstructured meshes which are used in practical engineering computations. 

\subsection{DNS of supersonic turbulent channel flow}

\begin{table}
 \centering
 \begin{tabular}{lccccccccccc}
 \hline
 Case & $M_b$ & $\Rey_b$ & $\Rey_{\tau}$ & $N_x$& $N_y$& $N_z$ &  $\Delta x^+$ & $\Delta y_w^+$ & $\Delta z^+$ &$C_f$ & $-B_q$\\
 \hline
 CH15-OF     & 1.5 & 6000 & 220 & 384 & 128 & 192 & 7.20 & 0.40  & 4.80 & 0.0078 & 0.049 \\
 CH15-FD     & 1.5 & 6000 & 220 & 256 & 128 & 192 & 10.8 & 0.70  & 4.80 & 0.0077 & 0.048 \\
 \hline
 \end{tabular}
 \caption{Flow parameters for DNS of plane channel flow for \textit{rhoEnergyFoam} in Mode A (CH15-OF)
  and for finite-difference solver~\citep{modesti_16} (CH15-FD).
   $M_b=u_b/c_w$ and $\Rey_b=2h\rho_bu_b/\mu_w$ are the bulk Mach and Reynolds number, respectively.
The computational box is $4\pi h\times2h\times4/3\pi$, with $h$ the channel half-height.
$N_i$ are the mesh points in each coordinate direction, and
$\Delta y_w^+$ is the distance of the first grid point from the wall, 
$\Delta x^+$, $\Delta z^+$ are the streamwise and spanwise grid spacings, in wall units.
$B_{q}=q_w/(\rho_w c_p u_{\tau}T_w)$ is the heat flux coefficient and $C_f=2\tau_w/(\rho_bu_b)$
is the skin friction coefficient.}
\label{tab:channel}
\end{table}
 
\begin{figure}
 \begin{center}
  (a)
  \includegraphics[width=5.0cm]{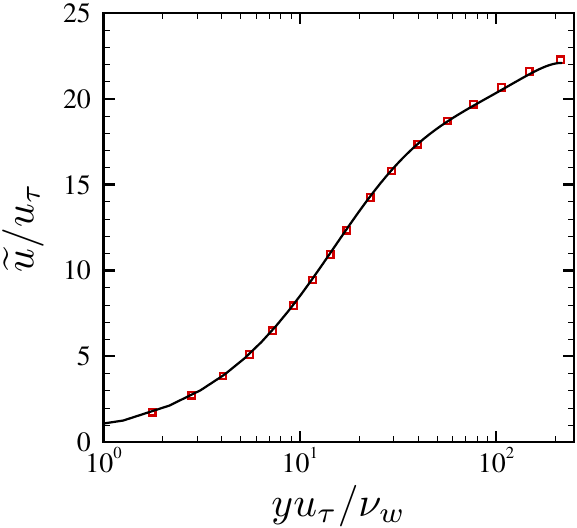} \hskip1.em
  (b)
  \includegraphics[width=5.0cm]{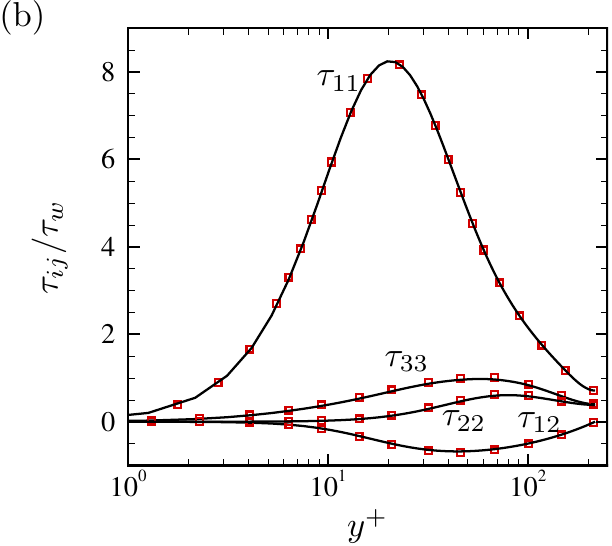}\\ \vskip1.em
  \caption{DNS of turbulent flow in plane channel: distribution of mean velocity (a) and Reynolds stresses (b) in wall 
          units for CH15-OF (solid lines) and CH15-FD (squares), for the
          DNS listed in Tab.~\ref{tab:channel}.}
  \label{fig:channel}
 \end{center}
\end{figure}

In order to test \textit{rhoEnergyFoam} for fully resolved compressible turbulent flows we carry out
DNS of supersonic channel flow at bulk Mach number $M_b=u_b/c_w=1.5$, and bulk Reynolds number
$\Rey_b = 2 h \rho_b u_b / \mu_w=6000$, where $u_b$ and $\rho_b$ are the bulk channel velocity and
density, $c_w$ is the speed of sound evaluated 
at the wall, and $h$ is the channel-half width. Supersonic channel flow is 
a common prototype of compressible wall-bounded turbulence, and several
database have been developed, spanning a wide range of Reynolds numbers~\citep{coleman_95,lechner_01,modesti_16}. 
In this flow case a Cartesian mesh is used fine enough that no artificial diffusion is needed, 
hence the solver is operated in Mode A.
The results obtained with \textit{rhoEnergyFoam} are compared with DNS data
obtained with a finite-difference sixth-order accurate energy-preserving solver~\citep{modesti_16} 
(see Tab.~\ref{tab:channel}).
Figure~\ref{fig:channel} compares the mean velocity and
the Reynolds stresses distributions in wall units and Favre density scaling (denoted with tildas), 
namely friction velocity $u_{\tau}=(\tau_w/\rho_w)^{1/2}$, and viscous length scale $\delta_v=\nu_w/u_{\tau}$.
The excellent agreement provides convincing evidence for the effectiveness of the
solver for DNS of compressible turbulent flows.

\subsection{RANS and DES of flow past circular cylinder}

\begin{table}
 \centering
 \begin{tabular}{lccccc}
 \hline
 Case & $M_\infty$  & $C_D$ &$-C_{pbase}$ & $\mathrm{St}_0$  & ${\Delta t}_{av}u_\infty/D$\\ 
 \hline
 URANS     & 0.1  & 0.28 &0.35  & -   & 220  \\
 DES       & 0.1  & 0.35 &0.44  & 0.31 & 150 \\
 URANS~\citep{catalano_03}     &  -  &  0.40 &0.41  & 0.31& 200 \\
 LES~\citep{catalano_03}       &  -  & 0.31 &0.32  & 0.35& 200 \\
 Exp.~\citep{shih_93}&  -   & 0.24 &0.33  & 0.22& - \\
 \hline
 \end{tabular}
 \caption{Main estimated properties for turbulent flow around circular cylinder.
         URANS and DES are carried out using \textit{rhoEnergyFoam} in Mode B,
         and compared with previous numerical simulations and experimental data.
         $C_D$ and $C_{pbase}$ are the drag coefficient and the base pressure 
         coefficient, respectively, 
         $\mathrm{St_0 = f_0 D / u_0}$ is the typical Strouhal number,
         and ${\Delta t}_{av}$ is the time averaging interval.}
\label{tab:cylinder}
\end{table}
 
\begin{figure}
 \begin{center}
  \includegraphics[width=6.0cm]{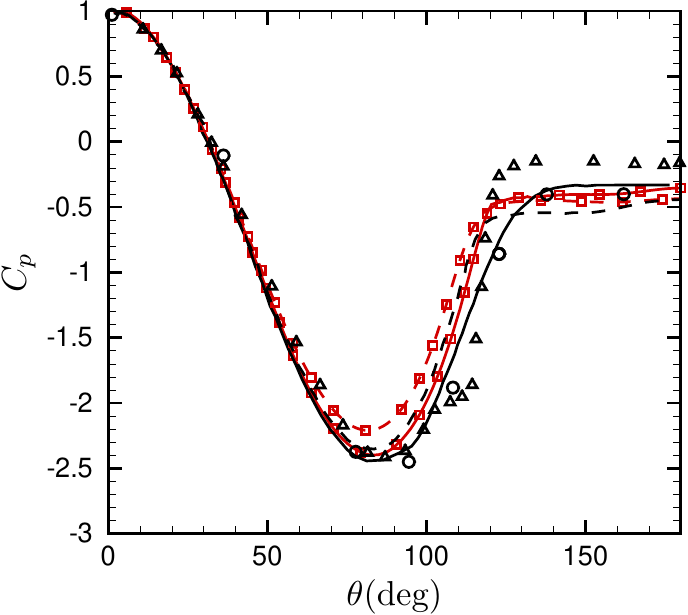}
  \caption{Numerical simulation of flow around circular cylinder:
           wall pressure coefficient obtained from 
           \textit{rhoEnergyFoam} in Mode B with URANS (solid line and squares) and
           DES (dashed line with squares), compared with URANS (solid) and
           LES (dashed) by~\citet{catalano_03} and with experiments 
           by~\citet{warschauer_71} (triangles) and \citet{zdravkovich_97}
           (circles).
         } 
  \label{fig:cp_cyli}
 \end{center}
\end{figure}
\begin{figure}
 \begin{center}
  (a)
  \includegraphics[width=5.0cm]{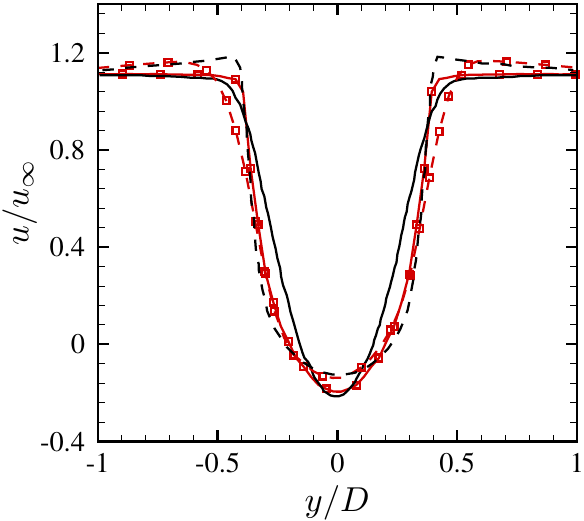} \hskip1.em
  (b)
  \includegraphics[width=5.0cm]{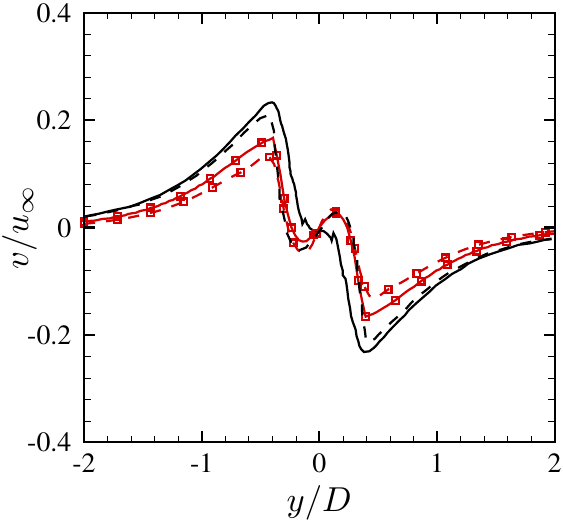} \\ \vskip1.em
  (a)
  \includegraphics[width=5.0cm]{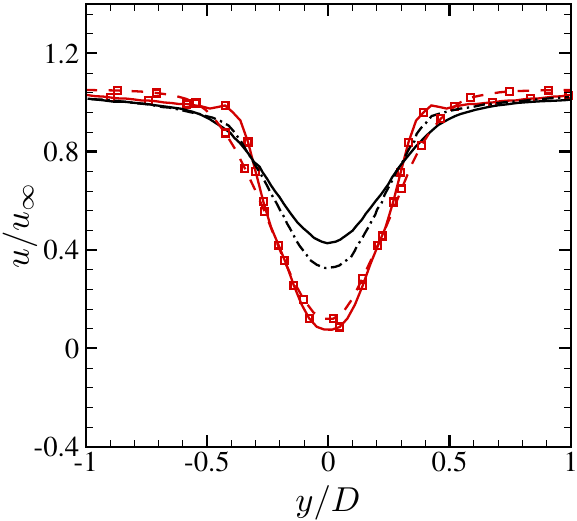} \hskip1.em
  (b)
  \includegraphics[width=5.0cm]{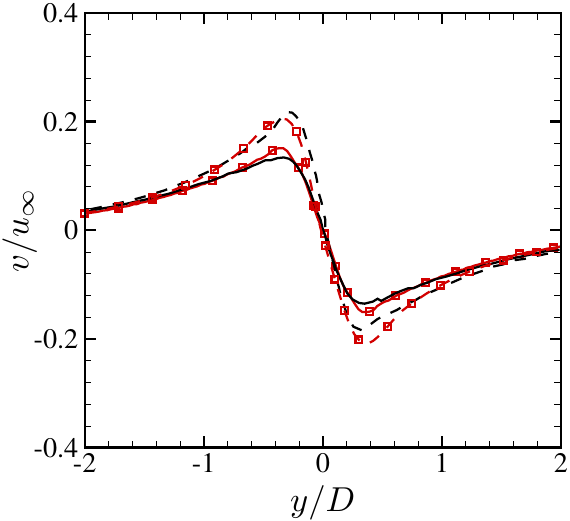} \\ \vskip1.em
  \caption{Numerical simulation of flow around circular cylinder:
           mean velocity profiles at $x/D=0.75$ (a)-(b) and $x/D=1.5$ (c)-(d)
           for URANS (solid line with squares) and
           DES (dashed line with squares) compared with URANS (solid) and
           LES (dashed) by~\citet{catalano_03}.
         } 
  \label{fig:vel_cyli}
 \end{center}
\end{figure}
 
The turbulent flow around a circular cylinder is here numerically studied by means of
\textit{rhoEnergyFoam} in mode B, through both
unsteady Reynolds-averaged Navier-Stokes simulation (URANS) and detached-eddy simulation (DES),
relying on the classical Spalart-Allmaras turbulence model~\citep{spalart_92}
and its DES extension~\citep{spalart_97}, respectively.
The free stream Mach number is $M_\infty=u_\infty/c_\infty=0.1$, where
$u_\infty$ and $c_\infty$ are the free stream velocity and speed of sound,
and the Reynolds based on the cylinder diameter is $\Rey_D = \rho_\infty u_\infty D / \mu_w$, with
$\rho_\infty$ the free stream density and $\mu_w$ the wall viscosity.
An O-type mesh is used for DES with $N_r \times N_\theta \times N_z = 256 \times 256 \times48$ cells 
in a $L_r \times L_z = 20 D \times 2 D$ domain,
whereas the same mesh with $N_z=1$ is used for URANS.
The mesh is stretched towards the cylinder with the first off-wall mesh point 
at $y^+\approx 150-200$, hence we rely on the use of wall functions for proper wall-treatment~\citep{piomelli_02}. 
Specifically, Spalding's equilibrium law-of-the-wall is used~\citep{spalding_61}.
Isothermal no-slip boundary conditions are imposed at the wall, whereas
inlet/outlet boundary conditions are used for all variables at the far field,
with the turbulent viscosity set to ${\mu_t}_0=3\mu_w$.

Table~\ref{tab:cylinder} shows the flow parameters used for the simulations, as well as the main flow properties
including the drag and the base pressure coefficient, and the typical nondimensional frequency in the cylinder wake, as
estimated from analysis of the pressure time spectra.
The numerical results are compared with previous numerical simulations~\citep{catalano_03} and experiments~\citep{shih_93}.
The main difference with respect to those is the absence of sensible vortex shedding in the present URANS, which is probably to be traced 
to the use of wall functions. Shedding is observed in DES, with global flow parameters in reasonable agreement
with other sources.
The wall pressure coefficient and the mean velocity profiles in the cylinder wake are further scrutinized 
in Figs.~\ref{fig:cp_cyli}, \ref{fig:vel_cyli}.
Comparison is overall satisfactory for both the pressure coefficient and the velocity profiles,
with the main difference that a longer cylinder wake is observed both in URANS and DES with respect to 
the reference numerical simulations of~\citet{catalano_03}.
Again, this deviation may be ascribed to imprecise prediction of the separation point 
caused by approximate wall treatment.

\subsection{Supersonic flow over forward-facing step}

\begin{figure}
 \begin{center}
  \includegraphics[width=10cm]{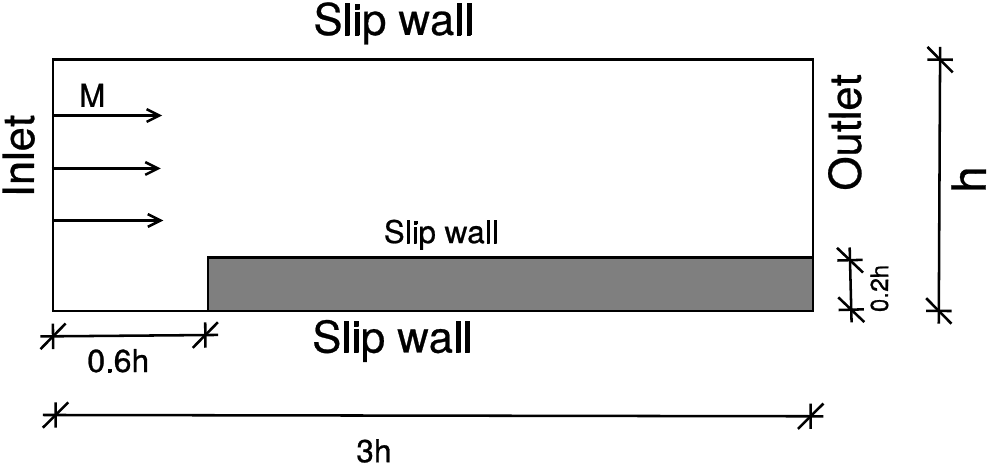}
  \caption{Sketch of the computational setup for flow over a forward-facing step~\citep{woodward_84}.
         } 
  \label{fig:step_sketch}
 \end{center}
\end{figure}

\begin{figure}
 \begin{center}
  \hspace{3.3cm}\raggedright{(a)}\\
  \hspace{3.7cm}\includegraphics[height=3.6cm]{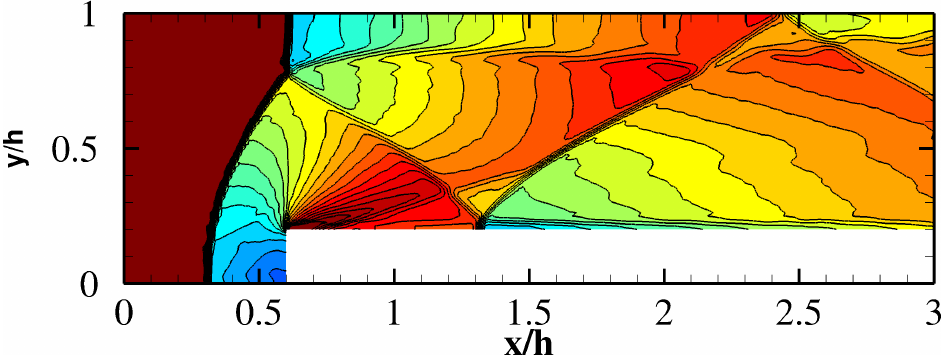}  \\
  \hspace{3.3cm}\raggedright{(b)}\\
  \hspace{3.7cm}\includegraphics[height=3.6cm]{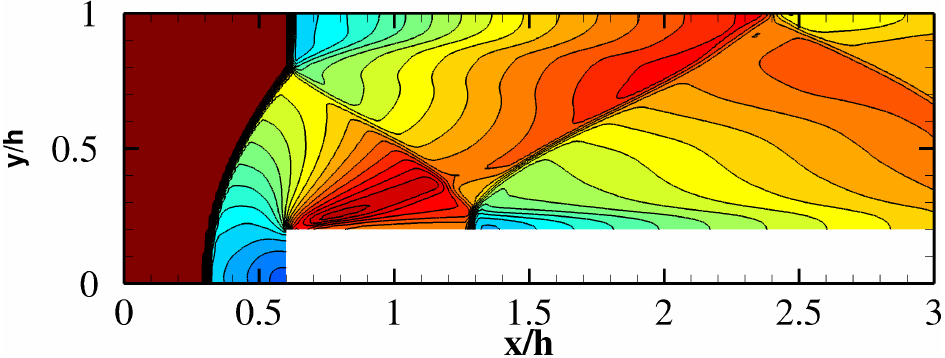} \\ \vskip1.em
  \hspace{3.3cm}\raggedright{(c)}\\
  \hspace{4.2cm}\includegraphics[width=9.3cm,]{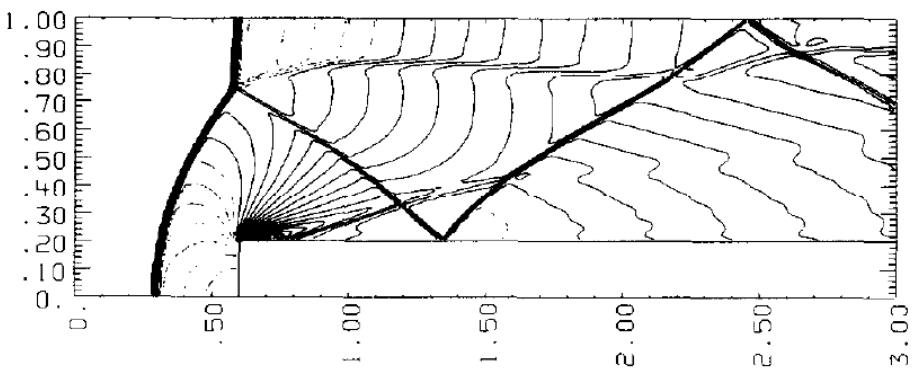} \\ \vskip1.em
  \caption{Supersonic flow past forward-facing step at
           $M_\infty=3$. 30 Mach number contours are shown in the
           range $-0.92 \le \rho \le 2.86$ (color scale from blue to red) for \textit{rhoEnergyFoam} (a),
           \textit{rhoCentralFoam} (b) and \citet{woodward_84} (c).}
  \label{fig:step_mach}
 \end{center}
\end{figure}

\begin{figure}
 \begin{center}
  \includegraphics[height=3.6cm]{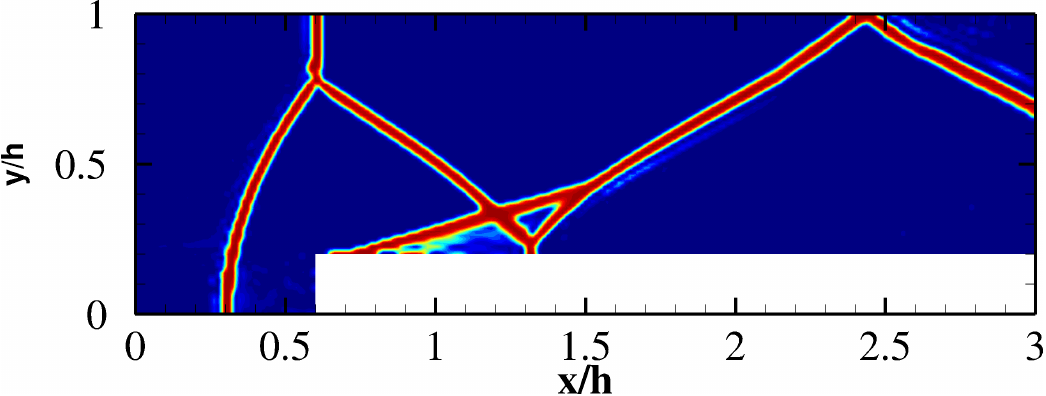}
  \caption{Supersonic flow past forward-facing step at
           $M_\infty=3$: contours of shock sensor, as defined in Eqn.~\ref{eq:sensor}.
           24 levels are shown in the range $0.05 \le \theta \le1$, 
           corresponding to the flow field shown in Fig.~\ref{fig:step_mach}(a).
         } 
  \label{fig:step_ducros}
 \end{center}
\end{figure}

The study of the inviscid flow over a forward-facing step was originally 
proposed by~\citet{emery_68} to compare
shock-capturing schemes. In particular, we consider 
the flow configuration used by~\citet{woodward_84},
in which the supersonic flow in a channel at $M_\infty=3$ 
faces a step of height $0.2h$, where $h$ is the channel height.
The total length of the channel is $3h$, the step leading edge is at $0.6h$
from the inlet and the mesh is uniform, with 
$N_x\times N_y=240\times80$ cells in the coordinate directions 
(see Fig.~\ref{fig:step_sketch}). Slip boundary conditions
are imposed at the top and lower walls, and all variables
are extrapolated at the outlet.

For this test case the solver is run in Mode C, 
with threshold value of the shock sensor $\theta^*=0.05$. 
Fig.~\ref{fig:step_mach} shows Mach number contours 
for \textit{rhoEnergyFoam} and \textit{rhoCentralFoam} compared with the
reference solution from~\citet{woodward_84}.
Inspection of the shock pattern shows that, despite qualitative
similarities, \textit{rhoEnergyFoam} delivers additional flow details 
which are barely visible
with \textit{rhoCentralFoam}. In particular the slip line issuing from the
quadruple point near the top wall 
in Fig.~\ref{fig:step_mach} is evanescent in \textit{rhoCentralFoam},
because of its higher numerical diffusion. 
Quantitative differences are also found in the prediction of the Mach stem
at the step wall, which is much taller in \textit{rhoCentralFoam}.
Figure~\ref{fig:step_ducros} shows contours
of the shock sensor corresponding to the field
shown in Fig.~\ref{fig:step_mach}(a), which highlights regions
in which the convective diffusive flux is activated ($\theta \ge 0.05$).
This is a convincing confirmation that numerical diffusion 
is only activated in close vicinity of shocks.

\subsection{Transonic flow over the ONERA M6 wing}

\begin{figure}
 \begin{center}
  (a)
  \includegraphics[height=5cm]{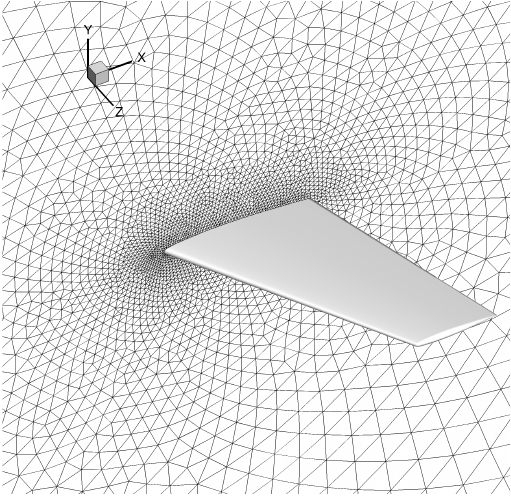} \hskip1.em
  (b)
  \includegraphics[height=5cm]{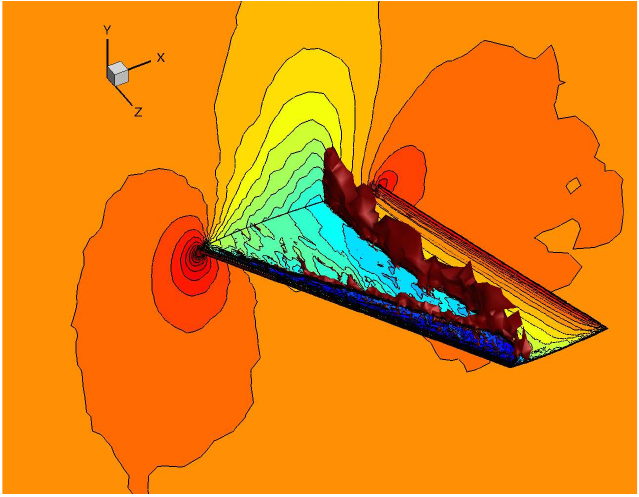}
  \caption{Unstructured mesh around the ONERA M6 wing (a),
          and computed pressure field with superposed iso-surface of shock
          sensor ($\theta=0.6$, in red) (b). 32 pressure contours are shown,
          in the range $0.2 \le p/p_\infty \le 1.3$ (color scale from blue to red).} 
  \label{fig:m6_mesh}
 \end{center}
\end{figure}
 
\begin{figure}
 \begin{center}
  (a)
  \includegraphics[width=5cm]{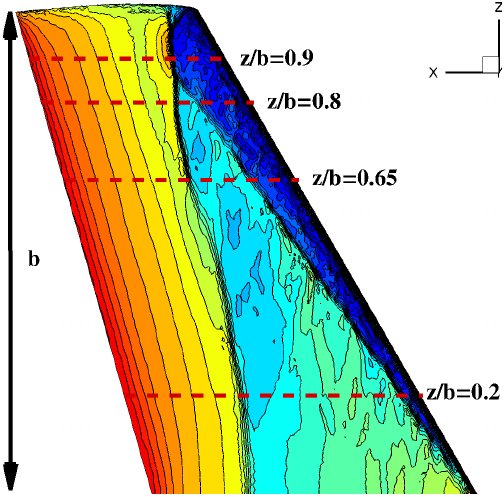} \hskip1.em
  (b)
  \includegraphics[width=5cm]{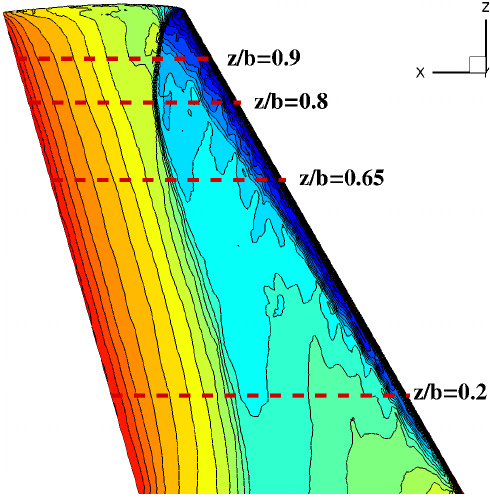}
  \caption{Flow around ONERA M6 wing: computed pressure contours on the wing surface for \textit{rhoEnergyFoam} (a) and \textit{rhoCentralFoam} (b).
           32 levels are shown in the range $0.3 \le p/p_\infty \le 1.3$ (color scale from blue to red).
           The dashed lines denote the wing sections used in Fig.~\ref{fig:m6_cp}.}
  \label{fig:m6_wing}
 \end{center}
\end{figure}
 
\begin{figure}
 \begin{center}
  (a)
  \includegraphics[width=4cm]{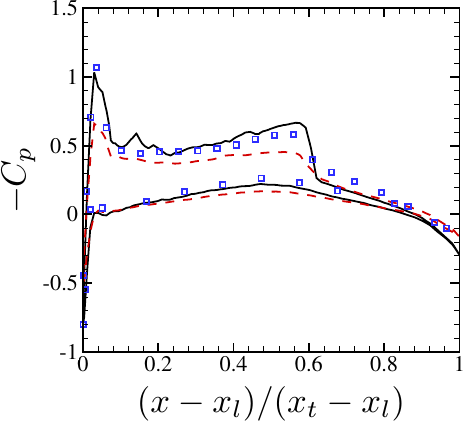} \hskip1.em
  (b)
  \includegraphics[width=4cm]{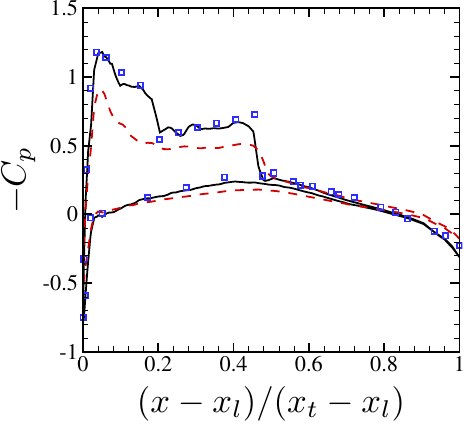}\\
  (c)
  \includegraphics[width=4cm]{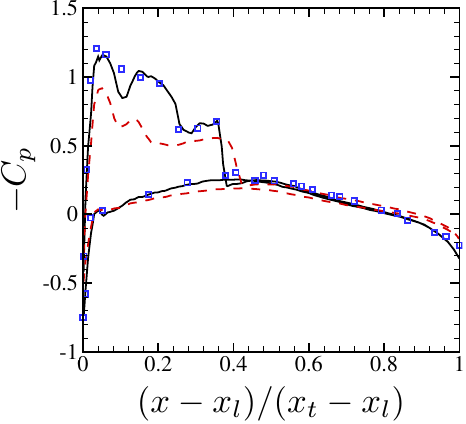} \hskip1.em
  (d)
  \includegraphics[width=4cm]{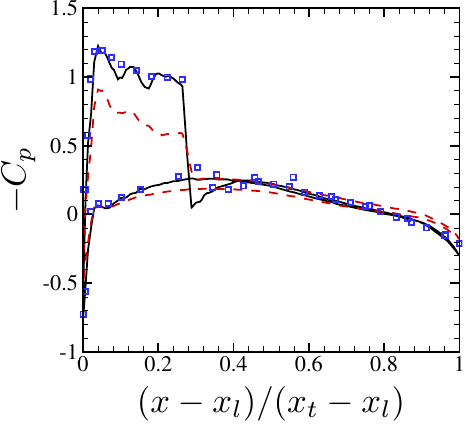} \\ \vskip1.em
  \caption{Flow around ONERA M6 wing: pressure coefficient ($C_p=(p-p_\infty)/(1/2\rho_\infty u_\infty^2)$) 
    at various wing sections (see Fig.~\ref{fig:m6_wing}): (a) $z/b=0.2$,
   (b) $z/b=0.65$, (c) $z/b=0.8$, (d) $z/b=0.9$, for \textit{rhoEnergyFoam} (solid lines)
   \textit{rhoCentralFoam} (dashed lines) and experimental data~\citep{schmitt_79} (square symbols).
   $x_l$ and $x_t$ denote the coordinates of the leading edge and trailing edge of each wing section, respectively.}
  \label{fig:m6_cp}
 \end{center}
\end{figure}

Results of numerical simulations of the inviscid flow past
the ONERA M6 wing~\citep{schmitt_79} are reported here, at free stream Mach number $M_\infty=0.8395$,
and angle of attack $\alpha=3.06^{\circ}$. An unstructured mesh including 341797 tetrahedral cells
is used (see Fig.~\ref{fig:m6_mesh}), within an outer computational box of size
$L_x \times L_y \times L_z = 10c \times10c \times5c$, where $c$ is the 
chord at the wing root section. 
Numerical simulations have been carried out using both \textit{rhoCentralFoam} and \textit{rhoEnergyFoam} in Mode C,
and compared with experimental data.
Figure~\ref{fig:m6_mesh}b shows the pressure field computed with \textit{rhoEnergyFoam}
with an overlaid iso-surface of the shock sensor, which highlights the presence of two 
shock waves, a primary one roughly at the middle of the wind chord, and
a secondary one close to the leading edge, eventually coalescing near the wing tip.

Figure~\ref{fig:m6_wing} shows the computed pressure field on the suction surface of the wing
for \textit{rhoEnergyFoam} (panel a) and \textit{rhoCentralFoam} (panel b), which highlights
qualitative differences between the two solvers. Although the main flow features are captured
by both the solvers, it seems that the leading-edge shock is much fainter in
\textit{rhoCentralFoam}, and the primary shock is much thicker especially towards the wing root,
owing to the diffusive nature of the solver.
A more quantitative evaluation is carried out in Fig.~\ref{fig:m6_cp}, where we compare
the computed distributions of the pressure coefficient with the experimental data of \citet{schmitt_79}, 
at the four wing sections indicated with dashed lines in Fig.~\ref{fig:m6_wing}.
At the innermost section (panel a) the primary shock is is rather weak,
and barely apparent in \textit{rhoCentralFoam}, whereas \textit{rhoEnergyFoam} yields
favourable prediction of both shock strength and position.
At intermediate sections (panels b,c) both shocks are present, which are again correctly captured
by \textit{rhoEnergyFoam}, whereas \textit{rhoCentralFoam} shows excessive smearing.
At the outermost section (panel d) the primary and the secondary shock merge
into a single stronger shock, whose amplitude is well captured by \textit{rhoEnergyFoam}.

\subsection{Transonic flow over the RAE-2822 airfoil}

\begin{table}
 \centering
 \begin{tabular}{lcc}
 \hline
 Case & $C_l$ & $C_d$  \\
 \hline
 \textit{rhoEnergyFoam}     & 0.713  & 0.0133 \\  
 \textit{rhoCentralFoam}    & 0.725  & 0.0185 \\
 Experiment~\citep{cook_77} & 0.743  & 0.0127 \\
 \hline
 \end{tabular}
 \caption{Lift and drag coefficient of RAE 2822 airfoil at $M_\infty = 0.725$, $\Rey_c = 6.5\times10^6$, 
         $\alpha=2.31^\circ$, as predicted by
         \textit{rhoEnergyFoam} and \textit{rhoCentralFoam}, compared with experimental
         data~\citep{cook_77}.}
\label{tab:rae}
\end{table}

\begin{figure}
 \begin{center}
  (a)
  \includegraphics[width=5cm]{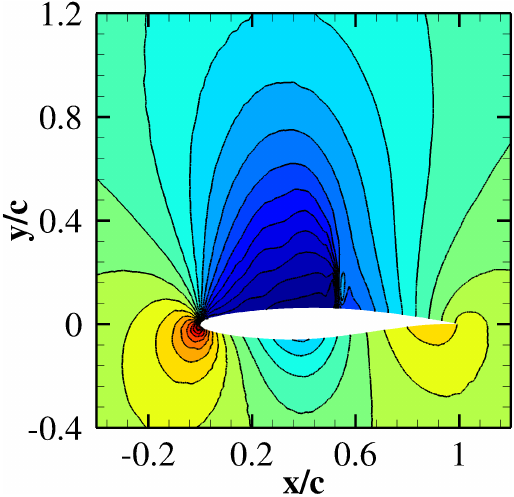} \hskip1.em 
  (b)
  \includegraphics[width=5cm]{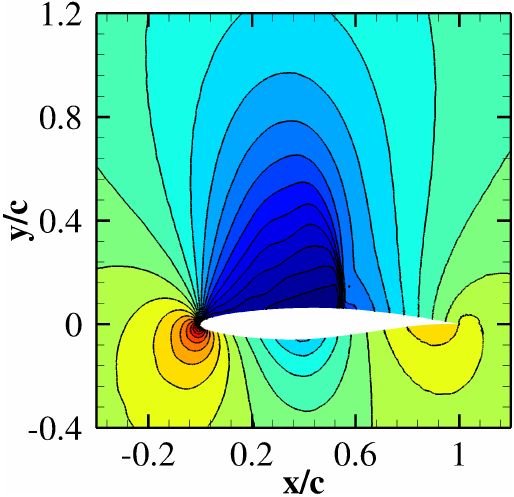}
  \vskip 1em
  \caption{RANS of flow over RAE 2822 airfoil: computed pressure field as predicted by \textit{rhoEnergyFoam} (a)
          and \textit{rhoCentralFoam} (b). 24 contour levels are shown in the range $0.6 \le p/p_\infty \le 1.4$, in color scale from blue to red.} 
  \label{fig:rae_pp}
 \end{center}
\end{figure}

\begin{figure}
 \begin{center}
  \includegraphics[width=6cm]{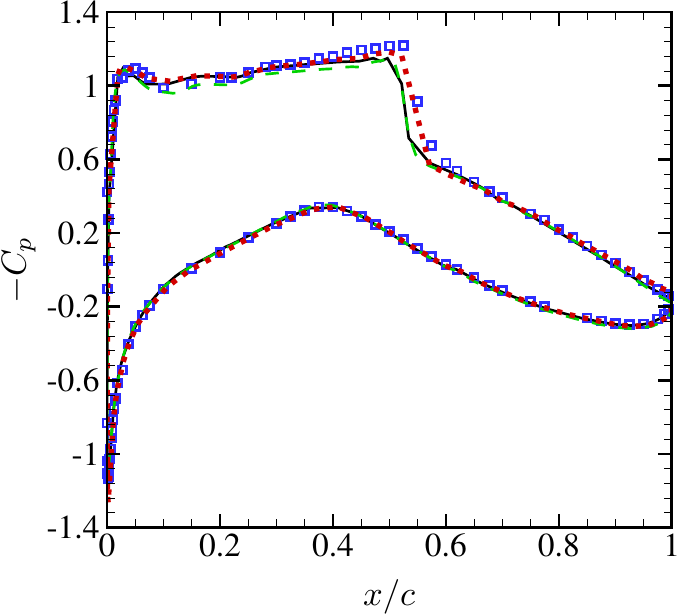}
  \caption{RANS of flow over RAE 2822 airfoil: pressure coefficient predicted by \textit{rhoEnergyFoam} (solid lines), \textit{rhoCentralFoam} (dots), compared with previous RANS~\citep{nelson_09} (dashed lines) and experimental data~\citep{cook_77} (square symbols).} 
  \label{fig:rae_cp}
 \end{center}
\end{figure}

The transonic flow past RAE 2822 airfoil~\citep{nelson_09} has been simulated through RANS,
using the standard Spalart-Allmaras model.
The flow conditions corresponds to those of test case 6 in the experiments of \citet{cook_77},
namely free stream Mach number $M_\infty=0.729$,
chord Reynolds number $\Rey_c=\rho_\infty u_\infty c/\mu_\infty=6.5\times10^6$, and angle of
attack $\alpha=2.31^\circ$.
A C-type structured mesh is used which includes $369 \times 256$ cells.
The far field boundary is at approximately
$20$ chords from the wall, where inlet/outlet boundary conditions
are enforced, whereas isothermal no-slip boundary conditions are imposed at the airfoil wall.
The distance of the first mesh point off the wall ranges between $\Delta_y^+=25-180$, hence the wall
is modeled through Spalding's wall function.
Table~\ref{tab:rae} shows the lift and drag coefficient predicted by \textit{rhoEnergyFoam} in Mode C
and \textit{rhoCentralFoam}, as compared with experimental data~\citep{cook_77}.
The agreement is quite good, with some overestimation of drag from \textit{rhoCentralFoam}.
The computed pressure fields are compared in Fig.~\ref{fig:rae_pp}, which shows
the presence of a single normal shock on the suction side, and very minor differences
between the two solvers.
Detailed comparison of the pressure coefficient with experiments~\citep{cook_77} 
and simulations~\citep{nelson_09}, shown in Fig.~\ref{fig:rae_cp}, is satisfactory
for both solvers, although in this case \textit{rhoCentralFoam} seems to be closer to experiments,
and \textit{rhoEnergyFoam} closer to previous simulations.

\section{Conclusions} \label{sec:conclusions}

A novel numerical strategy has been proposed for accurate simulation of smooth and shocked compressible flows
in the context of industrial applications.
The algorithm relies on the use of an underlying energy-consistent, non-diffusive numerical scheme,
which is locally augmented with the diffusive numerical flux of the AUSM scheme, in an amount 
dependent on the local smoothness of the flow on the computational mesh.
Three modes of solver operation have been suggested, based on the intent of the simulation.
We have found that fully resolved simulations (i.e. DNS) can be handled with no numerical diffusion (Mode A).
Smooth unresolved flows (i.e. DES and RANS) require some small amount of numerical diffusion,
granted by the pressure diffusive flux of AUSM (Mode B).
Shocked flows require further addition of the convective diffusive flux of AUSM for stability (Mode C).
For the sake of showing simplicity and generality of the approach, the method has been implemented in the OpenFOAM$^{\textregistered}$ library. 
A broad range of academic-to-applicative test cases have been presented to highlight the main features of the solver.
The simulation of homogeneous isotropic turbulence and Taylor-Green flow show that the solver operated in Mode A is capable of discretely 
preserving the discrete total kinetic energy from convection in the inviscid limit, whereas the baseline version of the OpenFOAM$^{\textregistered}$
solvers herein tested cannot. This features, besides being essential for DNS, is also appealing for URANS and DES. The applicative test cases 
here presented in fact support the statement that the use of low-diffusive numerics yields better representation of the flow physics, 
in contrast to highly diffusive schemes which tends to blur many features of the flow field.
This is reflected in improved quantitative prediction of local and global force coefficients in
applied aerodynamics test cases.

{\bf Acknowledgements}\\
We acknowledge that the numerical simulations reported in this paper have been carried out on the Galileo cluster based at CINECA, Casalecchio di Reno, Italy, using resources from the SHAPE project.

\section{Appendix}\label{sec:appendix}

Referring to Fig.~\ref{fig:cell}, the AUSM convective and pressure flux to be used in Eqn.~\eqref{eq:num_flux} are given
below, based on the AUSM$^+$-up formulation~\citep{liou_06}
\begin{equation}
\hat{\mathbf{f}}_{ON}^{AUSM} = -\frac{c_{ON}}{2}\left[\left( \frac{1}{2}\delta m_{ON}-|M_{ON}|\right)\boldsymbol{\varphi}_L
 +
\left( \frac{1}{2}\delta m_{ON}+|M_{ON}|\right)\boldsymbol{\varphi}_R
 \right],
\end{equation}
\begin{equation}
 \hat{\mathbf{p}}_{ON}^{AUSM} = -\frac{1}{2} \delta \mathbf{p}_{ON}, 
\end{equation}
\begin{equation}
 M_{ON} = \frac{M_R+M_L}{2} -\frac{1}{2}\delta m_{ON} + M_p,
\end{equation}
\begin{equation}
\delta m_{ON}=\left[\Delta \mathcal{M}(M_R)-\Delta \mathcal{M}(M_L)\right],\quad
\Delta\mathcal{M}(M)=\mathcal{M}_{(4)}^+(M)-\mathcal{M}_{(4)}^-(M),
\end{equation}
\begin{equation}
M_p=- \frac{k_p}{f_a}\max{(1-\sigma \overline{M}^2,0)}\,\frac{2\,(p_R-p_L)}{(\rho_L+\rho_R)c_{ON}^2}.
\end{equation}
The speed of sound at the cell interface is evaluated as $c_{ON} = (c_L+c_R)/2$ and
$\overline{M}^2=\left({u_n}_L^2+{u_n}_R^2\right)/(2 c_{ON}^2)$,
$M_0^2=\min{(1,\max{(\overline{M}^2,M_\infty^2)})}$,
$f_a(M_0)=M_0(2-M_0)$,
with $k_p=0.25$, $k_u=0.75$, $\sigma=1$.
The diffusive pressure flux is given by
\begin{equation}
\delta p_{ON}=\left[p_R\Delta\mathcal{P}(M_R)
 - p_L\Delta\mathcal{P}(M_L)\right]-2 M_u,\quad
\Delta\mathcal{P}(M) = \mathcal{P}_{(5)}^+(M) - \mathcal{P}_{(5)}^-(M),
\end{equation}
where
\begin{equation}
M_u = -\frac{1}{2}k_u\mathcal{P}^+_{(5)}(M_L)\mathcal{P}^-_{(5)}(M_R)
\left(\rho_L+\rho_R\right)\left(f_a c_{ON}\right)\left({u_n}_R-{u_n}_L\right),
\end{equation}
The subscript $L,R$ refers to the two sides of the cell interface, which have
have been reconstructed through the Minmod limiter, also available in the OpenFOAM$^{\textregistered}$ library.
We further define the split Mach numbers $\mathcal{M}_{(m)}$ as $m$-th degree polynomials 
\begin{equation}
\mathcal{M}^{\pm}_{(1)}(M)=\frac{1}{2}\left(M\pm|M|\right),
\end{equation}
\begin{equation}
\mathcal{M}^{\pm}_{(2)}(M)=\pm\frac{1}{4}\left(M\pm1\right)^2,
\end{equation}
\begin{equation}
\mathcal{M}^{\pm}_{(4)}(M)=
\begin{cases}
\mathcal{M}^{\pm}_{(1)}(M)&if \,\,\,|M|\ge1\\
\mathcal{M}_{(2)}^{\pm}(M)\left(1\mp16\beta\mathcal{M}^{\mp}_{(1)}(M)\right)&if \,\,\,|M|< 1.
\end{cases}
\end{equation}
$\mathcal{P}^{\pm}_{(5)}$ is also defined in terms of the split Mach numbers, as follows
\begin{equation}
\mathcal{P}^{\pm}_{(5)}(M)=
\begin{cases}
\frac{1}{M}\mathcal{M}_{(1)}^{\pm}(M)&if \quad|M|\ge1\\
\mathcal{M}_{(2)}^{\pm}(M)
\left[\left(2\pm-M\right)\mp16\alpha M 
\mathcal{M}_{(2)}^\mp(M)\mathcal{M}_{(1)}^\pm(M)\right]&if \quad|M|< 1 .
\end{cases}
\end{equation}
Following~\citet{liou_06}, we set $\alpha = 3/16 (-4 + 5 f_a^2)$, $\beta=1/8$.
\bibliographystyle{model1-num-names}
\bibliography{references} 
\end{document}